\documentclass[letterpaper, 10 pt, conference]{ieeeconf}  % Comment this line out
\IEEEoverridecommandlockouts                              % This command is only
\usepackage{graphics} % for pdf, bitmapped graphics files
\usepackage{graphicx}
\usepackage{float}
\usepackage{booktabs}
\usepackage{placeins}
\usepackage{epsfig} % for postscript graphics files
\usepackage{amsmath} % assumes amsmath package installed
\usepackage{amssymb}  % assumes amsmath package installed
\usepackage{tikz}
\usepackage{standalone}
\usepackage{bondgraphs}
\usepackage{url}
\usepackage{blox}
\usepackage{gensymb}
\usepackage{breakurl}
\usepackage{booktabs}
\usepackage{tabularx}
\usepackage{hyperref}
\usepackage{booktabs,amsfonts,dcolumn}
\usetikzlibrary{shapes,arrows}
\usetikzlibrary{calc, patterns, decorations.pathmorphing, decorations.markings, shapes.geometric, arrows, positioning, angles}

\mathchardef\mhyphen="2D
\mathchardef\mslash="202F

\newcommand{\RN}[1]{%
  \textup{\uppercase\expandafter{\romannumeral#1}}%
}

\allowdisplaybreaks
\mathchardef\mhyphen="2D
\mathchardef\mslash="202F

\title{\LARGE \bf
Adaptive Compliance Shaping with Human Impedance Estimation}
\author{Huang Huang$^{1}$, Henry F. Cappel$^{2}$, Gray C. Thomas$^{1,3}$, Binghan He$^{1}$ and Luis Sentis$^{2}$% <-this % stops a space
\thanks{The authors are with the $^{1}$Department of Mechanical Engineering and $^{2}$Department of Aerospace Engineering in the University of Texas at Austin}%
\thanks{$^{3}$The author is supported by a NASA Space Technology Research Fellowship (NSTRF) NNX15AQ33H.}%
\thanks{Send correspondence to $\;${\tt\small huangh at utexas dot edu}.}
}

\usepackage{dcolumn}
\usepackage{multirow}

\newcommand\copyrighttext{%
  \scriptsize 
  Accepted for publication in American Control Conference (ACC)
  \textcopyright 2020 IEEE. Personal use of this material is permitted. Permission from IEEE must be obtained for all other uses, in any current or future media, including reprinting/republishing this material for advertising or promotional purposes, creating new collective works, for resale or redistribution to servers or lists, or reuse of any copyrighted component of this work in other works.
  DOI: \href{https://ieeexplore.ieee.org/document/9147875}{10.23919/ACC45564.2020.9147875}
  }
\newcommand\copyrightnotice{%
\begin{tikzpicture}[remember picture,overlay]
\node[anchor=south,yshift=10pt] at (current page.south)
{\fbox{\parbox{\dimexpr\textwidth-\fboxsep-\fboxrule\relax}{\copyrighttext}}};
\end{tikzpicture}%
}

\begin{document}

\newcolumntype{L}[1]{>{\raggedright\arraybackslash}p{#1}}
\newcolumntype{C}[1]{>{\centering\arraybackslash}p{#1}}
\newcolumntype{R}[1]{>{\raggedleft\arraybackslash}p{#1}}

\maketitle
\thispagestyle{empty}
\pagestyle{empty}
\copyrightnotice

\vspace{-10pt}
\begin{abstract}
Human impedance parameters play a key part in the stability of strength amplification exoskeletons. While many methods exist to estimate the stiffness of human muscles offline, online estimation has the potential to radically improve the performance of strength amplification controllers by reducing conservatism in the controller tuning. We propose an amplification controller with online-adapted exoskeleton compliance that takes advantage of a novel, online human stiffness estimator based on surface electromyography (sEMG) sensors and stretch sensors connected to the forearm and upper arm of the human. These sensor signals and exoskeleton position and velocity are fed into a random forest regression model that we train to predict human stiffness, with a training set that involves both movement and intentional muscle co-contraction. Ground truth stiffness is based on system identification in essentially perturburator-style experiments. Our estimator's accuracy is verified both by the offline validation results and by the stability of the controller even as stiffness changes (a scenario where the ground truth stiffness is not available). Online estimation of stiffness is shown to improve the bandwidth of strength amplification while remaining robustly stable.
\end{abstract}

\section{Introduction}
	Robotic exoskeletons have been used for a range of applications including assistance with muscle impairment due to disease \cite{HaribHereidAgrawalGurrietFinetBoerisDuburocqMungaiMasselinAmesSreenathGrizzle2018CSM,farris2013preliminary,kwa2009development}, control mechanisms for tele-operation robots \cite{huang2017coordination,bergamasco1994arm}, and a means to augment the strength or increase the endurance of the human operator \cite{thomas2019compliance,he2019modeling,LeeKimBakerLongKaravasMenardGalianaWalshJNR2018,FontanaVertechyMarcheschiSalsedoBergamasco2014RAM}. Some researchers improve the performance of exoskeletons through feedback control \cite{HaribHereidAgrawalGurrietFinetBoerisDuburocqMungaiMasselinAmesSreenathGrizzle2018CSM} or offline and online optimization of control parameters \cite{LeeKimBakerLongKaravasMenardGalianaWalshJNR2018,ZhangFiersWitteJacksonPoggenseeAtkesonCollins2017Science}. This paper aims to improve the performance of a strength amplification exoskeleton---one that feedback couples exoskeleton joint torque to human joint torque in order to amplify human strength.
% Specifically, strength augmentation exoskeletons can assist workers repetitively lifting heavy payloads. 
% many types of exoskeletons that are not necessarily force amplification
	
	The stability of force amplification exoskeletons, like impedance controlled robots for physical human robot interaction, depends on the human impedance---and the exoskeleton must guarantee this coupled stability despite the variability in the human's behavior. Medically oriented studies often model the human as a spring, mass, damper system with time-varying parameters \cite{CANNON1982111,Bennett1992}. A more conservative model---where the human is a passive system---can provide very strong coupled stability guarantees \cite{thomas2019compliance,adams1999stable}, however this wider space of possible human models restricts controller performance  \cite{ArvidKeeminkHermanKooijArnoStienen2018}. An estimate of human stiffness with lower uncertainty has the potential to improve bandwidth for both human-robot interaction controllers and amplification exoskeletons \cite{thomas2019compliance,BuergerHogan2007,tsumugiwa2002variable}. 
% 	\cite{thomas2019compliance,he2019modeling,he2019complex} have shown the potential performance improvement for the controller by using a more accurate bound of human stiffness. 
	
	Exoskeletons can accomplish strength amplification through various control frameworks including adaptive control \cite{chen2016adaptive}, admittance control\cite{lecours2012variable}, impedance control \cite{karavas2015tele}, loop-shaping design with a bounded human impedance \cite{he2019modeling}, and by independently shaping the human and exoskeleton side compliance \cite{thomas2019compliance}. Ref.~\cite{he2019modeling} emphasizes remaining robustly stable and used system identification with the human in the loop in order to obtain a robust model of a SISO ``amplification plant''. In this framework it is clear how widening the uncertainty restricts the choice of crossover point and closed loop bandwidth. The framework in \cite{thomas2019compliance} emphasizes what dynamics behaviors are possible with the exoskeleton by specifying behavior in terms of two dynamic compliance transfer functions (exo-side and human-side). This framing makes it easy to design the controller to avoid instability with different human stiffnesses. A physical spring in \cite{thomas2019compliance} guaranteed a minimum compliance for the spring and human system and was used to design the controller. But an online estimate of human stiffness could provide the same information, without softening the human's connection to the exoskeleton.
	
	A common approach to measuring human stiffness is to impose a perturbation torque and measure deflection \cite{Mussa-Ivaldi2732}. However, this method is only effective offline \cite{kim2009estimation,mobasser2006method,shin2009myokinetic}.
% 	Human impedance properties are difficult to estimate online due to their dynamic nature and lack of direct measurement \cite{kim2009estimation,mobasser2006method,shin2009myokinetic}. 
    Online stiffness estimation methods include biological models \cite{shin2009myokinetic,franklin2003estimation,pfeifer2012model} as well as artificial neural networks \cite{kim2009estimation,mobasser2006method}, with only a subset of the estimation methods generalizing to multiple subjects \cite{shin2009myokinetic,pfeifer2012model}.   
	Most studies focusing on stiffness estimation use sEMG sensors  \cite{kim2009estimation,mobasser2006method,shin2009myokinetic,franklin2003estimation,pfeifer2012model}, but physical deflection sensors may offer a less noisy means to gain information from the human \cite{han2013active}. Our approach similarly combines sEMG with low cost stretch sensors (deflection-varying resistors). Although many studies have successfully estimated human impedance parameters, few have applied them to exoskeletons.

	Research has been done to incorporate other human property estimates into controllers \cite{guiliuzhang2019,NgeoTameiShibataOrlandoBeheraSaxenaDutta2013,kawaseHiroyukiKoike2012}, with most studies focusing on the estimation of applied torque or human intention \cite{karavas2015tele,LiWangSunYangXieZhang2014,ChengHuangHuang2013,KiguchiHayashi2012}. In many cases this torque estimation is used as an alternative to contact force sensors between the human and exoskeleton. The researchers in \cite{kawaseHiroyukiKoike2012} perform a dynamic stiffness estimation using a musculoskeletal model for a power assist exoskeleton, but focus on the reduction of vibrations due to EMG noise.

	In this paper we apply online estimation of human stiffness to adapt the force feedback gains of a strength amplification exoskeleton according to the estimated human stiffness. Our online human stiffness estimator uses a novel combination of sensors, and arguably improves over the state of the art for estimating the stiffness of the human elbow, boasting an R factor of 0.993 (c.f. 0.9266 in \cite{kim2009estimation}), and a 17 $\mathrm{Nm/rad}$ max error (c.f. 30 $\mathrm{Nm/rad}$ in \cite{mobasser2006method} and 80 $\mathrm{Nm/rad}$ in \cite{shin2009myokinetic}). We also contribute a novel controller adaptation scheme (based on the compliance shaping framework \cite{thomas2019compliance}) that uses bounded-error stiffness information to improve bandwidth while remaining stable. This controller is then experimentally validated to A) remain stable as stiffness changes, B) lose stability when fed incorrect stiffness information, and C) improve strength amplification bandwidth relative to a robust control design.
% 	This paper proposes an accurate online estimation of stiffness using a random forest predictor taking biological data from stretch sensors, and sEMG sensors. Our training data set is extended to include stiffness calculated from voluntary movement. The approach is validated by online shaping of a controller to improve both bandwidth and amplification. This paper contributes in two ways. First, we demonstrate the accuracy of a dual biological sensor approach to online stiffness estimation. Second, we show that this strategy can effectively be incorporated into a controller to increase both the bandwidth and amplification ultimately leading to an improved performance for the user. 

\section{Online Stiffness Estimation}
We first propose an approach to estimate human stiffness online by using a trained random forest model taking advantage of signals from sEMG and stretch sensors as well as exoskeleton velocity and position.
\subsection{Apparatus}
\begin{figure}
    \rule{0pt}{8pt}\\
    \centering{
    \resizebox{\columnwidth}{!}{
    \def\svgwidth{1.4\columnwidth}
    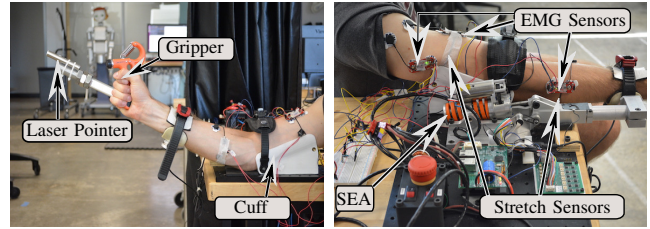}}
    \caption{The P0 exoskeleton (Apptronik Systems Inc., Austin, TX) with an ATI Mini40 (ATI Industrial Automation, Apex, NC) force sensitive cuff located near the middle of the forearm. The subject holds a grip-strength exercise device to modulate co-contraction in the muscles at the elbow. The subject is instrumented with 3 sEMG sensors and 2 stretch sensors that are used to estimate stiffness.}
    \label{fig:hardware}
\end{figure}

    We use a single degree of freedom elbow joint exoskeleton for this research. The P0 exoskeleton (Apptronik Systems Inc., Austin, TX), as shown in Fig. \ref{fig:hardware}, is a 3 bar linkage device powered by a series elastic actuator (SEA) with a spring force tracking bandwidth of 10 $\mathrm{Hz}$ and reliable actuator torque conversion using a linkage table. The exoskeleton includes a 6-axis force torque sensor measuring the human exoskeleton contact forces. The human rests his or her upper arm on a white 3D printed mount beside the actuator. Exoskeleton position $\theta$ is measured by an encoder at the joint and contact torque $\tau_c$ is measured by the force torque sensor. The moment of inertia of the exoskeleton is 0.1 $\mathrm{kg\cdot m^2}$ without any additional weight, but provides the option to include additional external weights. A laser pointer is attached to the end of the long bar to assist with precise position movement projecting onto a white board one meter in front of the subject wearing the exoskeleton. The white board contains three lines referring to initial position and upper and lower bounds of movement. A deviation around $\pm 3 \degree$ from those lines is acceptable.
    
    In addition, we utilize 3 Myowear sEMG sensors (SparkFun Electronics, Niwot, CO) located on the upper  arm  and  forearm  (biceps  brachii,  triceps brachii,  and  brachioradialis  muscles) of the subject and 2 stretch sensors (Images Scientific Instruments Inc., Staten Island, NY) attached around the middle of the forearm and upper arm connected to an Arduino Mega 2560 (SparkFun) by a breadboard. The sampling frequency for all sensors is 250 $\mathrm{Hz}$. The full setup of the apparatus including the exoskeleton and the peripheral sensors are shown in Fig.~\ref{fig:hardware}.

\subsection{Experimental Protocol}
The experimental protocol was approved by the Institutional Review Board (IRB) at the University of Texas at Austin. One healthy, male subject wore the 3 sEMG sensors and 2 stretch sensors during the experiments.

The experiments are divided into 2 sections. The first consists of 11 experiments in which the participant maintains a constant equilibrium position while the exoskeleton imposes a torque comprising a piece-wise constant bias and a sinusoidal excitation with constant frequency and amplitude. In order to obtain reference signal values for all sensors, the participant initially holds a constant posture for 20 seconds, aligning the laser pointer to a target. The first 20 seconds includes gravity compensation, with no bias torque. Following this procedure, the exoskeleton induces bias forces ranging from 0 $\mathrm{Nm}$ to 9.5 $\mathrm{Nm}$ in 0.5 $\mathrm{Nm}$ steps occurring in 3 second intervals. Because we noticed there tends to be larger errors for the low bias torques, we repeated the first five bias forces twice. The participant is asked to maintain the same constant position and apply no voluntary compensation torque. Movement is induced by the sinusoidal signal, which has a constant frequency of 1 $\mathrm{Hz}$ and amplitude of 1.5 $\mathrm{Nm}$. This experiment is repeated 11 times (denoted $\RN{1}$.1-11), with a 30 second resting period between every five bias force transitions as well as a minimum of 2 minutes resting period between each of the 11 experiments. In $\RN{1}$.1, the subject holds nothing. To induce muscle co-contraction, $\RN{1}$.2-11 introduce a hand-grip exercise tool with an adjustable load. The participant squeezes a gripper beginning with 22 $\mathrm{lb}$ for the second trial and up to 82 $\mathrm{lb}$ for the final trial. 

The second set of experiments maintains the same procedure as the first experiment set except the participant voluntarily moves his or her arm at 0.5 $\mathrm{Hz}$, using three optical targets for the midpoint and two extremes of the oscillation. In this experiment the sinusoidal excitation has a constant frequency of 1.7 $\mathrm{Hz}$ and an amplitude of 2.5 $\mathrm{Nm}$. The bias force increases from 0 $\mathrm{Nm}$ to 8 $\mathrm{Nm}$ in step of 2 $\mathrm{\mathrm{Nm}}$ occurring in 15 second intervals. All other parameters and procedures remain consistent with the first set of experiments (including the variation of grip strength). This set of experiments is denoted as $\RN{2}$.1-11.

\setlength{\belowcaptionskip}{-4pt}

\begin{figure}[!tbp]
    \centering
    \def\svgwidth{1.0\linewidth}
    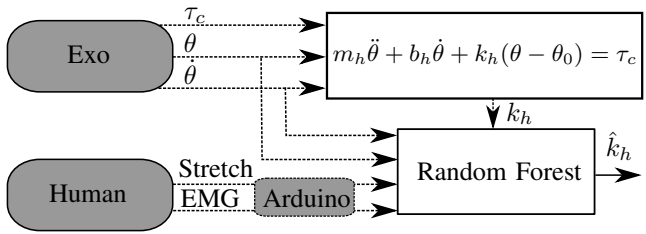
    \caption{Diagram of training scheme for random forest predictor. Stiffness $k_h$ is estimated using least squares fitting in the time domain, and is used as the ground truth for training the stiffness predicting random forest.}
    \label{fig:forest_diagram}
\end{figure}
\subsection{Methods}

\begin{figure*}[h!]
\centering
    \def\svgwidth{1.0\linewidth}
    \rule{0pt}{8pt}\\
    \scalebox{0.95}{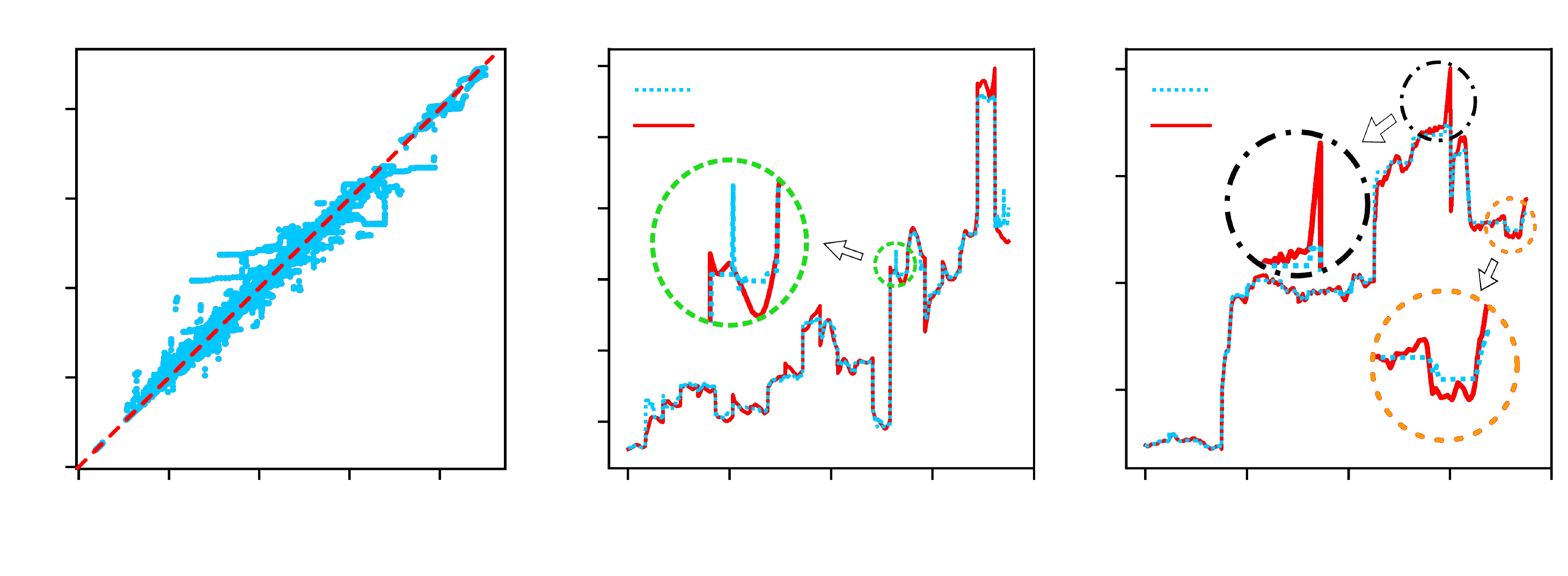}
\caption{Random forest predictor results. $\hat{k}_h$ is the estimated stiffness from our random forest predictor and $k_h$ is the reference stiffness calculated from the time domain regression. Fig.~\ref{fig:estimation_results}(a) shows the linear relationship between the estimated stiffness and the reference stiffness for all experiments $\RN{1}$.1-11 and $\RN{2}$.1-11. The blue dots are the data points and the red dash line is the reference line of $y=x$. Fig.~\ref{fig:estimation_results}(b) shows estimation results from experiment group I. Fig.~\ref{fig:estimation_results}(c) shows estimation results from experiment group II.}
\label{fig:estimation_results}
\end{figure*}

\setlength{\belowcaptionskip}{-9pt}

\subsubsection{Data Preprocessing}
In both experiment sections signals from 3 sEMG sensors are amplified, rectified, and integrated and then passed through a second order low pass filter with cutoff frequency of 60 $\mathrm{rad/s}$ and damping ratio of 0.707. We use the average signal values from 2 stretch sensors and 3 sEMG sensors in the first 20 seconds of each experiment as initial reference signal values for that experiment. These values are subtracted from the sEMG and stretch sensors' data to get the variation data for the 5 sensors. The absolute values of processed data from the stretch and sEMG sensors are denoted as $S1$, $S2$ and $E1$, $E2$, $E3$ respectively.

In the first experiment section, exoskeleton position and velocity, and contact torque are filtered with the same second order low pass filter to calculate the reference stiffness. 

In the second experiment section, we use a second order butterworth bandpass filter \cite{butter} with cutoff frequency of 1.2 $\mathrm{Hz}$ and 10 $\mathrm{Hz}$ for exoskeleton position and velocity, and contact torque to filter out the influence of human voluntary movement when calculating the reference stiffness. 

For both sections, exoskeleton position and velocity are filtered by the same second order low pass filter to build the training and validation data set.

\subsubsection{Time Domain Regression}
In order to obtain a reference stiffness value for training the online estimation model and validating the accuracy, we use a linear regression for the time domain data regarding the dynamic equation\footnote{Here, we use a linear damping model to estimate the human's stiffness, because of difficulties implementing hysteretic damping in the time domain regression. Hysteretic damping models are likely more accurate\cite{he2019complex}, and we use them for the stability analysis.}
\begin{flalign}
     m_h\ddot{\theta} + b_h\dot{\theta} + k_h(\theta - \theta_0) = \tau_c \label{time_domain_regression}
\end{flalign}
where $\tau_c$ is the contact torque between the human and exoskeleton, $m_h$, $b_h$, and $k_h$ are the inertia, linear damping, and stiffness of the human, $\theta$, $\dot{\theta}$ and $\ddot{\theta}$ are the joint position, velocity and acceleration of the human, and $\theta_0$ is the equilibrium angle of the human spring (i.e. the human's desired position). In the case of a rigid connection between the human and exoskeleton, the human's joint position, velocity and acceleration are equal to the corresponding measurable properties of the exoskeleton. Through a linear regression between $\tau_c$ and $[\theta, \, \dot{\theta}, \, -1]$ for the corresponding experimental data ($\ddot{\theta}$ is not included due to the amplified noise from the double differentiation on joint position), we find the human stiffness $k_h$ as the reference stiffness, linear damping $b_h$, and offset spring torque $\tau_0 = k_h\theta_0$. Each linear regression includes a moving window of 400 points in time.

\subsubsection{Random Forest Predictor}
We use a random forest predictor from scikit-learn package \cite{scikit-learn} in Python to estimate muscle stiffness based on a 7-dimensional training data set, which includes the absolute value of exoskeleton position and velocity, filtered by the second order low pass filter, and S1, S2, E1, E2 and E3. The reference stiffness values are used as a supervisory signal. The model is structured with an estimator number of 50 and a maximum depth of 10 for each estimator to avoid over-fitting. The predictor is trained offline with data from both the first and second experimental sections. The full diagram of the model training procedure is outlined in Fig.~\ref{fig:forest_diagram}.

\subsection{Results}
We obtain 76350 offline shuffled data points where 50900 are used for offline training and the remaining 25450 are used as an offline validation set. The estimation results for all data sets using the trained random forest predictor give us a maximum error of 16.58 $\mathrm{Nm/rad}$ and an error variance of 2.55 $\mathrm{Nm^2/rad^2}$. The results are shown in Fig.~\ref{fig:estimation_results}(a). Estimation results for the validation data set only have a maximum error of 14.51 $\mathrm{Nm/rad}$ and an error variance of 3.01 $\mathrm{Nm^2/rad^2}$. 
Representations of accurate estimation results are shown in Fig.~\ref{fig:estimation_results}(b) and Fig.~\ref{fig:estimation_results}(c) respectively.

The quality of our predictor is high relative to other published predictors of human stiffness using sEMG data. From Fig.~\ref{fig:estimation_results}(a) we notice a significant linear relationship between stiffness estimation and reference stiffness. Comparing the estimation results with other similar research, our R factor 0.993 points to a stronger correlation than the best result of elbow stiffness in \cite{kim2009estimation} of 0.9266 (\cite{kim2009estimation} uses an artificial neural network to estimate multi-joint stiffness, but we only compare the elbow joint stiffness results). Our stiffness ranges from 5 to more than 90 $\mathrm{Nm/rad}$ which is a more practical range compared with \cite{kim2009estimation}'s smaller range of 1 to 3 $\mathrm{Nm/rad}$. Our predictor has a maximum error less than 17 $\mathrm{Nm/rad}$ while Fig.5 in \cite{mobasser2006method} shows a maximum error greater than 30 $\mathrm{Nm/rad}$ and the results in \cite{shin2009myokinetic} show a maximum error greater than 80 $\mathrm{Nm/rad}$. However, \cite{shin2009myokinetic} uses a different definition of elbow stiffness and includes data for nine subjects, which may influence their estimation accuracy. 
% \cite{pfeifer2012model} uses a physical model to estimate human's knee stiffness for multiple subjects while we estimate human's elbow joint stiffness for a single subject, so it's hard to compare our results with theirs.  
In addition, all the experiments in \cite{kim2009estimation,mobasser2006method,shin2009myokinetic,pfeifer2012model} are done without the human's voluntary movement, which weakens the validation of their models. Stiffness estimation in the presence of voluntary motion introduces new challenges, because these voluntary movements can be confused with the human's response to the perturbation. Our bandpass filter helps to remove the influence of human voluntary motion in the estimation procedure (the human's voluntary motion is below the lower cutoff frequency), but does not completely eliminate this influence. This implies that the reference human stiffness is not entirely trustworthy for the second experiment set.

The error between estimated stiffness and reference stiffness may come from three sources: error caused by incorrect sensor data, error caused by the imperfect predictor, and error due to incorrect reference stiffness. The green circle of Fig.~\ref{fig:estimation_results}(b) demonstrates a sudden peak in the stiffness estimate, a peak which is not reflected in the smooth reference stiffness. This kind of instant peak may be caused by inaccurate sensor data corrupting the inputs to the stiffness predictor. An erroneous momentary sensor value may be due to buffer error or electrical noise, which will cause the predictor to return an incorrect estimation result. In Fig.~\ref{fig:estimation_results}(c), the error shown in the orange circle may be a pure inaccuracy from the predictor while the error in the black circle may be caused by the incorrect reference stiffness. Since $k_h$ in Fig~\ref{fig:estimation_results}(c) is acquired using a band pass filter, this unusual sudden increase and decrease of reference stiffness in the black circle can be explained by human motion being abrupt enough to enter the bandpass region of the filter. 

In general, our predictor gives an accurate stiffness estimation for both stiffness in isometric conditions and during voluntary movement. This random forest predictor can be used for online stiffness estimation. If we eliminate the data from the stretch sensors in the training data set, we notice a decrease of R factor from 0.993 to 0.987 and an increase of maximum error from 16.58 to 19.32 $\mathrm{Nm/rad}$. The error variance also increases from 2.55 to 5.13 $\mathrm{Nm^2/rad^2}$ validating the importance of including the data from the stretch sensors.
% (!!!!!!!!!!maybe we can add one more graph to do the comparison between with and without sensor's data)
\begin{figure}[!tbp]
\rule{0pt}{8pt}\\
\centering
    \scalebox{0.75}{
    \def\svgwidth{1.0\linewidth}
    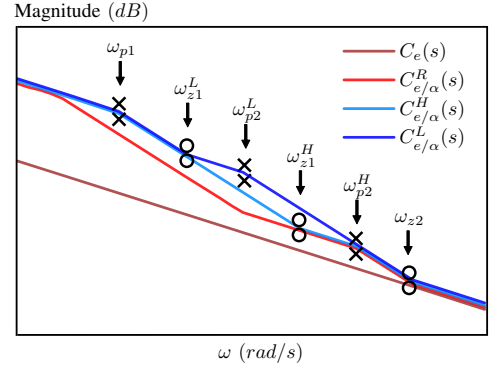}
    \caption{Conceptual bode plot shows the amplification performance for both the robust controller and the adaptive controller. $C_e(s)$ corresponds to the exoskeleton compliance. $C_{e \mslash \alpha}^H(s)$ and $C_{e \mslash \alpha}^L(s)$ correspond to the human side compliance of the exoskeleton using the adaptive controller when the human has a high stiffness and low stiffness. $C_{e \mslash \alpha}^R(s)$ corresponds to the human side exoskeleton compliance using the robust controller.}
    \label{fig:concept_bode}
\end{figure}
\begin{figure}[!tbp]
% \setlength{\belowcaptionskip}{-19pt}
% \small
\centering
    \rule{0pt}{8pt}\\
    \scalebox{0.90}{\def\svgwidth{1.0\linewidth}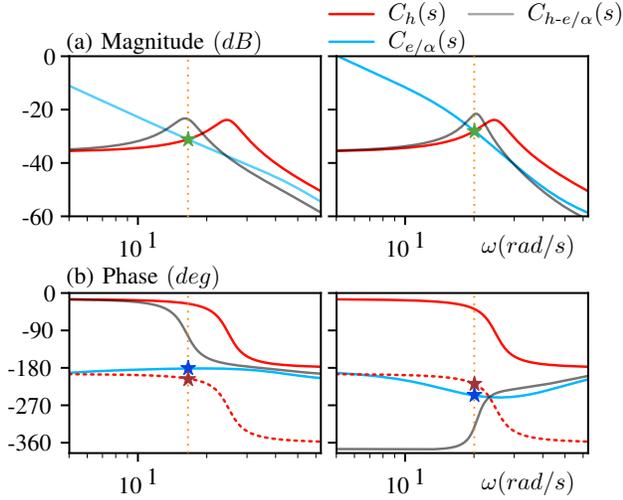}
    \caption{Bode plot showing stability behavior. The red dashed line in the phase plot is equal to $\phi(C_h(s))-180^{\circ}$. The phase difference between the blue line and red dashed line determines the stability of the system. The left graph shows a stable system and corresponding phase behavior of the human in exoskeleton with conservative values of $\lambda_1$ and $\lambda_2$. The right graph shows an unstable behavior corresponding to more aggressive values of $\lambda_1$ and $\lambda_2$. }
    \label{fig:bode_stability}
\end{figure}
\section{Application}
Since we have demonstrated that stiffness can be estimated online to a reasonable accuracy, we can now exploit this knowledge to design higher performance exoskeleton controllers.

\subsection{Controller Adaptation Scheme}

The relationship between exoskeleton position and external torque can be expressed as
\begin{equation}\label{compliance1}
   m_{e} s^2 \cdot \theta = \tau_e+\tau_c+\tau_s.
\end{equation}
where $\tau_e$ is environment torque, $\tau_c$ is torque applied by the human, and $\tau_s$ is our control input. Exoskeleton inertia $m_e$ includes the attached weight. We implement a compliance shaping amplification controller as $\tau_s = (\alpha(s) - 1) \tau_c$ so as to achieve the nominal behavior
\begin{equation}
    {m_{e} s^2} \cdot \theta = \tau_e + \alpha(s) \tau_c,
\end{equation}
where the human is amplified by a factor of $\alpha(s)$.
This choice of control does not alter the environment-side compliance of the exoskeleton, $C_e (s)=1/(m_{e} s^2)$. But it allows the human to feel an attenuated compliance $C_{e \mslash \alpha}(s)$ of the exoskeleton as 
\begin{flalign}
 C_{e \mslash \alpha} (s) = \frac{\alpha(s)}{m_{e} s^2},
\end{flalign}
which we refer to as the ``human-side'' compliance.

Our adaptation strategy determines a transfer function $\alpha(s)$ based on the measured human stiffness. We parameterize $\alpha(s)$ as
\begin{equation} \label{eq:alpha}
     \alpha (s) = \frac{{(s^2+2\zeta_0 \omega_{z1} s + \omega_{z1}^2)}{(s^2+2\zeta_1 \omega_{z2} s + \omega_{z2}^2)}}{{(s^2+2\zeta_0 \omega_{p1} s + \omega_{p1}^2)}{(s^2+2\zeta_0 \omega_{p2} s + \omega_{p2}^2)}}.
\end{equation}
The steady state amplification rate $\alpha_{ss} = (\omega_{z1}^2 \omega_{z2}^2)/(\omega_{p1}^2 \omega_{p2}^2)$. The amplification $\alpha(s)$ approaches unity at high frequencies, making the torque feedback ($1-\alpha(s)$) strictly causal, even though $\alpha(s)$ is not. For simplicity, we order the four natural frequency parameters $\omega_{p1},\ \omega_{z1},\ \omega_{p2},\ \omega_{z2}$ as shown in Fig.~\ref{fig:concept_bode}, and do not attempt to adapt the damping ratio $\zeta$ parameters. We place $\omega_{z2}$ at 10 $\mathrm{Hz}$ to avoid exceeding the bandwidth of the low level force controller, and this leaves us three free frequency parameters in the controller design. We remove one free parameter by fixing the desired steady state amplification ratio. As explained later, the gap between $\omega_{z1}$ and $\omega_{p2}$ must enclose a crossover frequency that depends on human stiffness. We constrain the remaining two degrees of freedom by choosing two tuning parameters $\lambda_1$ and $\lambda_2$ that ensure a sufficient distance between this crossover frequency and $\omega_{z1}$ and $\omega_{p2}$, 
\begin{flalign}
    \lambda_1 = \frac{\omega_{h \mhyphen e}}{\omega_{z1}}, \quad \lambda_2 = \frac{\omega_{p2}}{\omega_{h \mhyphen e}},
\end{flalign}
where $\omega_{h \mhyphen e} = \sqrt{k_h/m_{h \mhyphen e}}$ is the natural frequency of the human in the exoskeleton and $m_{h \mhyphen e}$ is the inertia of the human and exoskeleton including the attached weight. The forearm inertia $m_h$ has been measured for an average human at 0.1 kg m$^2$ in \cite{CANNON1982111}, but we do not know the inertia of our own subject.
% Since human's inertia is ignorable compared to the inertia of exoskeleton with attached weight, we have
% \begin{equation}
%     m_{h \mhyphen e}\approx m_e
% \end{equation}

Ultimately, we define our controller based on $\lambda_1$, $\lambda_2$, $\alpha_{ss}$ and the estimated value of $\hat{k}_h$:
\begin{flalign}
    \omega_{z1} &=\frac{\omega_{h \mhyphen e}}{\lambda_1}=\frac{1}{\lambda_1}\sqrt{\frac{\hat{k}_h}{m_{h \mhyphen e}}}\label{omega1},\\
    \omega_{p2} &= \lambda_2 \omega_{h \mhyphen e} = \lambda_2 \sqrt{\frac{\hat{k}_h}{m_{h \mhyphen e}}}\label{omega4},\\
    \omega_{p1} &= \frac{\omega_{z1} \omega_{z2}}{\sqrt{\alpha_{ss}} \cdot \omega_{p2}}\label{omega3}.
\end{flalign}
This allows us to change the shape of our amplification in real time. We refer to this real time compliance shaping controller as an adaptive controller in this paper. In contrast, without real time stiffness estimation, we have to use the most conservative bound of human stiffness to calculate $\omega_{z1}$, $\omega_{p1}$ and $\omega_{p2}$, which reduces our amplification bandwidth, $\omega_{p1}$. We refer to this as the robust controller.

\begin{figure}
    \centering{
\rule{0pt}{8pt}\\
    \resizebox{\columnwidth}{!}{
    \def\svgwidth{1.5\columnwidth}
    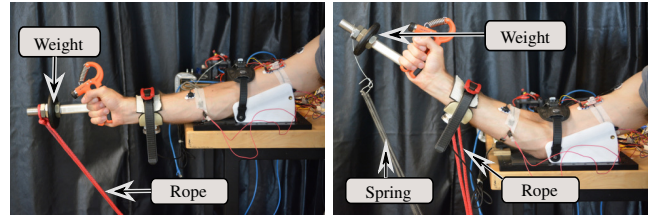}}
    \caption{Experimental setup to verify the improvement of the controller. The left picture shows the setup of the bandwidth test and the right picture shows the setup of the stability test. The rope is in place to maintain a constant position in the bandwidth test and limit the range of position to protect the actuator in the stability test. In both tests, a 1.25 $\mathrm{lb}$ weight is attached to the end of the long bar (though this has no effect on the bandwidth test where the output is locked). }
    \label{fig:demo_setup}
\end{figure}
The conceptual bode plot shown in Fig.~\ref{fig:concept_bode} illustrates the improved performance using stiffness estimation and shows the amplification performance in different frequencies and values of stiffness. It is straightforward to find a better amplification performance of the compliance shaping controller with online stiffness estimation because the amount of uncertainty handled by the controller is reduced. The difference between the lines corresponding to $C_{e \mslash \alpha}^H$ (the compliance shape when the human stiffness is high) and $C_{e \mslash \alpha}^L$ (the shape when it is low) indicates the controller's shape changing with different stiffness values. In either case the steady state amplification behavior continues until $\omega_{p1}$, a far higher bandwidth than that achieved by $C_{e \mslash \alpha}^R$, the compliance shape that is robust to both human stiffness extremes.

The stability analysis for these controllers is based on the complex stiffness model of human impedance proposed in \cite{he2019complex}, with
\begin{equation}
    C_h (s) = \frac{1}{m_h s^2+k_h + c_h j},
\end{equation}
where $c_h$ is the hysteretic damping of the human. According to \cite{he2019complex},
\begin{equation}
    \zeta_h = \frac{c_h}{2k_h} ,
\end{equation}
where $\zeta_h$ is the damping ratio of the human's elbow joint---which has been found to be nearly constant for repeated measurements of a subject \cite{he2019complex,MilnerCloutier1998EBR,LacquanitiLicataSoechting1982BC}. We use a conservative, constant damping ratio of 0.13 to represent our subject. 

The parallel connection between human compliance and human side exoskeleton compliance results in the total compliance of the human in the exoskeleton $C_{h \mhyphen e \mslash \alpha} (s)$ being a harmonic sum
\begin{equation}
    C_{h \mhyphen e \mslash \alpha} (s) = \left(\frac{1}{C_h (s)} + \frac{1}{C_{e \mslash \alpha}(s)}\right) ^ {-1}.
\end{equation}

% \begin{figure}[!tbp]
%     \input{boder.pdf_tex}
%     \caption{Boder plot}
%     \label{fig:concept_bode}
% \end{figure}

The stability of this system is determined by the phase margin of $\frac{C_{e \mslash \alpha}(s)}{C_h(s)}$.
 \begin{gather}
\frac{C_{e \mslash \alpha}(s)}{C_h(s)}=\frac{\alpha(s)}{m_e s^2}(m_h s^2+k_h+c_hj) 
 \end{gather}
Therefore, the stability of this system can also be determined by the ``human phase margin'' of $C_{e\mslash\alpha}(s)$, \begin{equation}
     \Delta \phi = \phi(C_{e \mslash \alpha}(s))-(\phi(C_h(s))-180^{\circ})\label{eq:phase_margin}.
\end{equation}
%  \Delta \phi &=  180^{\circ} + (\phi(C_{e \mslash \alpha}(s))-\phi(C_h(s)))\\
The two bode plots in Fig.~\ref{fig:bode_stability} show how large values of $\lambda_1$ and $\lambda_2$ produce a stable system (left) and how small values degrade the human phase margin and result in an unstable system (right). Note that the unstable system has a phase that rises rather than falling at the pole-pair---this indicates the poles are in the RHP.
% The margins in Fig.~\ref{fig:bode_stability}.(a) are so large that the amplification behavior is hard to notice in the magnitude plot, as most of it occurs below the plotted frequency range.

As mentioned before, we do not know the inertia of our subject. Fortunately, in \eqref{eq:phase_margin} reducing the phase of the human compliance increases the phase margin, and thus approximating human inertia as zero is conservative. We therefore choose values for $\lambda_1$ and $\lambda_2$ which guarantee stability for zero human inertia. In a more realistic test with human inertia based on \cite{CANNON1982111}, these parameters are confirmed to be stable.

\subsection{Experiment Validation}
We performed three tests to verify the stability, and bandwidth increase of the compliance shaping controller that incorporates the online stiffness estimation, as well as the significance of accurate online stiffness estimation. 

\subsubsection{Stability Test}
We verify stability of the two controllers using a step response test. The experimental apparatus shown in the right image of Fig. 4 incorporates a spring attached to the end of the exoskeleton to induce an external force on the device. The removal of this spring acts as a step force excitation to the system. 
% In addition, a 1.25 $\mathrm{lb}$ weight is attached to the end of the long bar.

The first experiment tests the robust controller. The participant wears the exoskeleton without the sEMG and stretch sensors and maintains a constant position while the spring is attached. After 10 seconds we remove the spring and observe the step response in the position signal. %Following another 10 seconds the spring is reattached and, again, the controller remains stable.
We repeat this procedure for a low stiffness (no gripper) and high stiffness case (the participant squeezes the gripper of 72 $\mathrm{lb}$).

\begin{figure}[!tbp]
\centering
\rule{0pt}{8pt}\\
    \def\svgwidth{0.45\textwidth}
    \scalebox{1.}{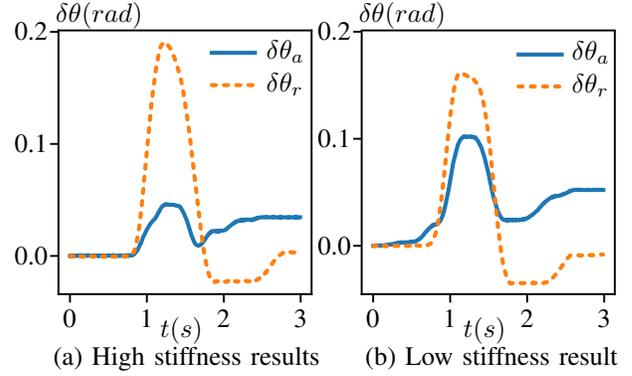}
    \caption{Stability test response shown by the exoskeleton position changing with time. $\delta\theta_a$ is the position change response of the adaptive controller and  $\delta\theta_r$ is the robust controller response.}
    \label{fig:stability}
\end{figure}
\begin{figure}[!tbp]\centering
\rule{0pt}{8pt}\\
\resizebox{0.93\columnwidth}{!}{
    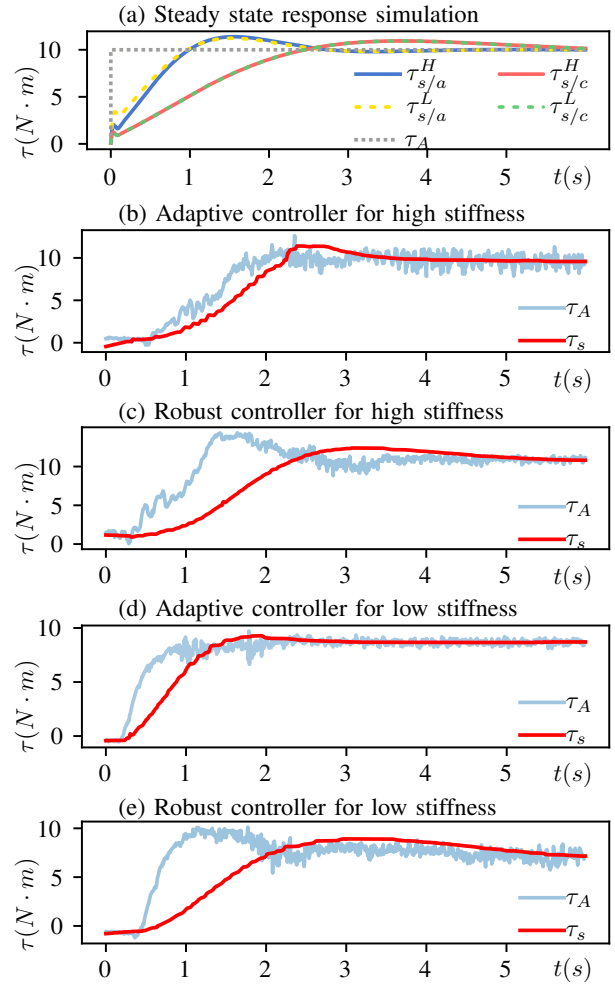}
    \caption{Steady state response for the bandwidth increase test. $\tau_s$ is the actuator torque. $\tau_A$ is equal to $-\alpha_{ss}\tau_c$ where $\tau_c$ is the contact force between the human and exoskeleton measured by the force sensor around the cuff. $\tau_A$ is the amplification torque we want to achieve. $\tau_{s\mslash a}^H$ and $\tau_{s\mslash a}^L$ are the simulated actuator torques of the adaptive controller in high stiffness and low stiffness. $\tau_{s\mslash c}^H$ and $\tau_{s\mslash c}^L$ are the simulated actuator torques of the robust controller in high stiffness and low stiffness.}
    \label{fig:bandwidth}
\end{figure}

For the second experiment we repeat the same procedure, but using the adaptive controller. The participant wears the sEMG sensors
% on three muscle groups (biceps brachii, triceps brachii, and brachioradialis muscles)
and stretch sensors to allow a real-time muscle stiffness estimate, which is also observed.
% positioned around the forearm and the midpoint of the biceps muscle on the upper arm. 

\subsubsection{Bandwidth Increase Test}
This experiment is designed to compare the bandwidth of the adaptive controller with the robust controller. The experimental setup shown in the left image of Fig.~\ref{fig:demo_setup} incorporates a rope attached to the end of the exoskeleton to maintain a constant position by pulling against the hard-stop. 

In order to verify the bandwidth improvement of the adaptive controller, the participant wears the exoskeleton and generates a (near) constant force for 10 seconds. Actuator torque is observed. This process is repeated for the robust controller. 
% , relaxes for 10 seconds, and again induces a constant force.
% For the purpose of maintaining a constant force during these trials a rope is attached to the end of the exoskeleton keeping the device in place.
\subsubsection{Instability Test}
The significance of accurate online stiffness estimation is measured by using the adaptive controller \emph{without} real stiffness estimate data. Instead, a dummy stiffness estimate (60 $\mathrm{Nm/rad}$) is used. In addition, the participant does not wear sEMG or stretch sensors. The setup is as the stability test, except that the step input is unnecessary. The subject maintains a constant position and relaxes their muscles for 10 seconds while the controller loses stability. After 10 seconds the participant maximally tenses their muscles and the controller regains stability. 

% We demonstrate the significance of including online estimation of stiffness by fixing a static stiffness value and showing stability only when the human's muscle stiffness matches this value.

% The controller is modified to rely only on a stiffness estimation near the peak of the subject's capability ($k_e = 60$ $Nm/rad$). In addition, the participant does not wear EMG or stretch sensors to estimate the stiffness value. The participant maintains a constant position and relaxes their muscles for 10 seconds while the controller loses stability. After 10 seconds the participant maximally tenses their muscles and the controller regains stability.
\subsection{Results}

\begin{figure}[!tbp]
\centering
\rule{0pt}{8pt}\\
\scalebox{.85}{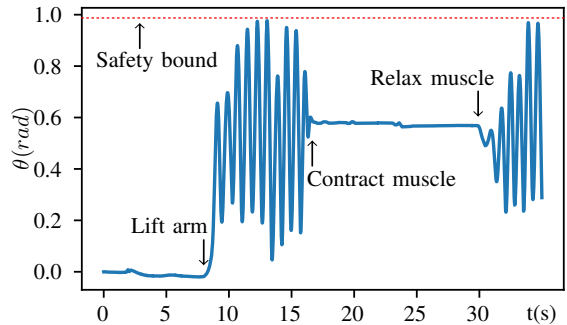}
    \caption{The instability test. The red dotted line at the top of the graph is the maximum position, as limited by the rope shown in the right picture of Fig.~\ref{fig:demo_setup}.}
    \label{fig:unstable}
\end{figure}

Results from these experiments are shown in Fig.~\ref{fig:stability}, Fig.~\ref{fig:bandwidth} and Fig.~\ref{fig:unstable} respectively.

Fig.~\ref{fig:stability} shows that both controllers give a stable response to an impulse input, however the adaptive controller produces a smaller vibration amplitude than the robust controller for both cases of high stiffness and low stiffness. The lower overshoot amplitude of the adaptive controller response may be due to a better human phase margin and correspondingly better damping ratio in the human--robot system.

Fig.~\ref{fig:bandwidth} shows both the simulation results of the steady state response with a step input (Fig.~\ref{fig:bandwidth}(a)) as well as experimental results (Fig.~\ref{fig:bandwidth}(b-e)). Fig.~\ref{fig:bandwidth}(b)(c) shows the comparison of the robust controller and the adaptive controller in the high stiffness case and Fig.~\ref{fig:bandwidth}(d)(e) shows the low stiffness case. The lag between $\tau_s$ and $\tau_A$ indicates the bandwidth of the controller. In both cases, the adaptive controller requires less time to achieve the target torque $\tau_A$ and therefore has a higher bandwidth. The experimental results appear consistent with the simulation results---large visual differences in the plots are largely due to the human input deviating from a perfect step.

Fig.~\ref{fig:unstable} shows the instability test result. When the adaptive controller has a discrepancy between the estimated stiffness value and the actual stiffness value, the system becomes unstable as shown in Fig.~\ref{fig:unstable}. This experiment highlights the importance of accurate stiffness estimation to our adaptive controller.

\section{Discussion}
Many studies performed on amplification exoskeletons have relied on conservative bounds of human impedance properties \cite{he2019modeling,BuergerHogan2007,he2019complex}. Due to the difficulties of online estimation of human muscle stiffness \cite{kim2009estimation,mobasser2006method,shin2009myokinetic,pfeifer2012model}, few studies have attempted to improve amplification controller performance using these properties.

In this paper, we propose an adaptive compliance shaping controller and demonstrate the improved performance due to stiffness estimation. The adaptive controller using the stiffness estimation provides increased stability and higher bandwidth than a comparable robust controller designed based on a conservative bound of human stiffness. We prove this improvement both theoretically and experimentally on a one DOF exoskeleton.

Accurate stiffness estimation is necessary to realize this compliance shaping controller. Our random forest predictor---using data from both sEMG and stretch sensors---was sufficiently accurate for this purpose. Our two experiment sections include training data from both isometric conditions and dynamic conditions with voluntary movement. Our estimation results appear to be more accurate than similar studies \cite{kim2009estimation,mobasser2006method}. The estimation results may be further improved with better and more reliable sensors, as well as by taking into consideration the time delay of the filter. A higher accuracy would allow us to use a lower safety bound $\lambda_1$ and $\lambda_2$ to achieve even higher bandwidth. 

In this paper, we only collected data from a single subject and trained a random forest model specified for this subject, which is non applicable to other subjects. In the future, we may include more subjects and train a more general random forest model applicable to multi subjects.

The convergence of the random forest predictor has not been proven, so it is difficult to make guarantees about the performance and safety of the predictor. As future work, we propose to integrate a backup safety controller \cite{thomas2018safety} to take over if the learning system fails. Such a backup controller could offer firm safety guarantees, but would not interfere with the controller if it was not misbehaving.

The bandwidth increase test and the stability test point to performance improvement that can be realized with information about human properties. In this paper, we use very conservative values of $\lambda_1$ and $\lambda_2$, calculated based on a zero human inertia assumption, for both the adaptive and robust controller, which limits the performance of both controllers. In future studies, we can use a more aggressive safety bound to achieve better performance for both controllers with accurate knowledge of human inertia. However, we can still expect the adaptive controller to outperform the robust controller. We believe this method can be applied to other kinds of controllers currently lacking knowledge of human impedance parameters. For instance the controllers in \cite{he2019modeling,BuergerHogan2007,he2019complex} may achieve similar bandwidth improvements with a similar system to update the human model online. 
\FloatBarrier
\bibliographystyle{IEEEtran}
\bibliography{main}
% \balance
\end{document}